# Twin Spotlight Beam Generation in Quadratic Crystals

**Raphaël Jauberteau**[1,2], **Sahar Wehbi**[1], **Tigran Mansuryan**[1], **Alessandro Tonello**[1], **Fabio Baronio**[2], **Katarzyna Krupa**[3], **Benjamin Wetzel**[1], **Stefan Wabnitz**[4], **and Vincent Couderc**[1,*]

[1] Université de Limoges, XLIM, UMR CNRS 7252, 123 Avenue A. Thomas, 87060 Limoges, France
[2] Dipartimento di Ingegneria dell'Informazione, Università di Brescia, via Branze 38, 25123, Brescia, Italy
[3] Institute of Physical Chemistry, Polish Academy of Sciences, Warsaw, Poland
[4] DIET, Sapienza University of Rome, Via Eudossiana 18, 00184 Rome, Italy

e-mail* vincent.couderc@xlim.fr

**Abstract**
Optical rogue waves have been extensively studied in the past two decades. However, observations of multidimensional extreme wave events remain surprisingly scarce. In this work we present the experimental demonstration of the spontaneous generation of spatially localized two-dimensional beams in a quadratic nonlinear crystal, which are composed by twin components at the fundamental and the second-harmonic frequencies. These localized spots of light emerge from a wide background beam, and eventually disappear as the laser beam intensity is progressively increased.

Solitons are self-sustained nonlinear waves, and the abundance of their observations in many different nonlinear media is a proof of their universal nature [1,2]. In fiber optics, ultrashort pulses usually spread in time as a result of group velocity dispersion (GVD). On the other hand, temporal solitons are special pulses which self-preserve their temporal envelope, when GVD is stably compensated by fiber nonlinearity. Solitons in optical fibers are exact solutions of the well-known 1-dimensional (1D) nonlinear Schrödinger equation (NLSE) [3-5]. Continuous waves or long pulse envelopes instead, can have a completely different fate: they can be unstable against temporal perturbations, and eventually break-up in a train of short pulses. Although the early stages of this process, which is known as modulational instability, can be described by a simple linear stability analysis, the nonlinear stage of the evolution of modulation instability can also be described in terms of analytical solutions of the NLSE. Among these nonlinear modulation waves, we may cite the so-called Akhmediev breathers [6], Peregrine solitons [7], and Kuznetsov-Ma breathers [8-9]. All of those nonlinear waves exhibit localization both in the temporal and in the longitudinal (or propagation) coordinates. Specifically, Peregrine solitons are often referred to as an example of waves that "appear from nowhere, and disappear without a trace" [10]. All of these waves have been clearly observed in optical fiber experiments, although most of their theory was first developed in the context of hydrodynamics [11-13]. In spite of being fully deterministic solutions of the 1D-NLSE, nonlinear modulation waves have been proposed as a fundamental model for rogue waves, which represent statistical extreme events in many different physical settings, such as oceanography. The randomness of the initial conditions for wave propagation associates rogue waves to the appearance of rare events, whose presence can be detected by long-tails in the probability distribution of, e.g., the wave height [14].

The origin of the success of the 1D-NLSE model is associated to its generality to describe wave propagation in weakly (cubic) nonlinear and dispersive media, as well as to its full integrability by the inverse scattering method. Notably, deterministic rogue waves have been also predicted in media with quadratic nonlinearity. In this case, the three-wave mixing equations apply, which is another example of fully integrable nonlinear wave model [15]. Incidentally, under the approximation known as cascading, the three-wave mixing equations can also be reduced to the 1D-NLSE.

If one replaces the temporal coordinate by a transverse spatial coordinate, the 1D-NLSE equation also describes beam propagation when diffraction (analogous to dispersion) is affecting the local beam size (analogous to the pulse temporal envelope). The analogy is not east do draw, since the transverse spatial domain is inherently bidimensional, while the time domain is described by a single dimension instead. A good approximation for a one-dimensional "flat" transverse space is the case of a slab waveguide (or possibly a very elliptical beam): spatial extreme waves have been predicted to appear in 1D spatial quadratic media [17-19]. A further extension could be considered for the case of an array of waveguides or multimode propagation in a waveguide: in both cases, multimodality can be seen as a synthetic transverse dimension. Solitons have been predicted in arrays of waveguides and in multimode fibers [20-21].

When the full bidimensional nature of the transverse spatial domain is considered, the beam propagation problem is far more complex, and many dynamical systems are no longer integrable. Nevertheless, both "solitons" (under their less demanding definition of nonlinear self-preserving waves) and modulation instabilities have been reported for these systems [22-26]. When the interacting optical waves have different linear group velocities in the presence of diffraction and nonlinearity, the nonlinear wave system increases in complexity. In nonlinear media with dominant quadratic susceptibility and in the presence of crystal anisotropy, spatial 2D solitons are conditionally stable solutions. This is the case of walking solitons [27], and related beam instabilities [28-29].

The possible existence of extreme waves in 2D wave systems is an open question, and a new territory which is largely yet to be explored. A first endeavor in this research direction was the numerical prediction of rogue waves, as reported in Ref. [30]. Spatial 2D rogue waves and caustics generated by nonlinear instabilities have been studied by means of a statistical approach in [31]. Some mathematical models also predict the existence of 2D deterministic rogue waves: for example, the three-wave mixing model which describes nonlinear coupling between a fundamental wave (FF) and its second harmonic (SH). Remarkably, under some approximations [32] one obtains an integrable model, in the frame of which the existence of extreme waves has been recently predicted [32]. Another important consideration is that, when dealing with the propagation of nearly plane waves or relatively wide beams, that is of intrinsically two-dimensional waves, the observation of spatial modulation instability in quadratic media has been reported [34-36]. However, when taking into account the presence of the temporal dimension, then the analysis of modulation instabilities should be extended to the case of 3D waves. In analogy with the 1D case [11], it is then natural to speculate whether in a 2D crystal with a quadratic nonlinearity, or even in 3D quadratic systems (when time is also considered), the decay (or the instabilities) of spatially localized solutions which is observed when, e.g., the input power is varied, can be interpreted as a signature of the underlying existence of rogue waves.

The present work demonstrates, we believe for the first time, the generation of extreme light beams in anisotropic crystals with quadratic nonlinearity. Specifically, we report the experimental observation of a deterministic 2D spatial twin spotlight beam (TSB), which appears from nowhere and disappears without a trace, as the input laser power is varied. By twin beam we mean a beam composed by two frequency components, one at the fundamental frequency (FF) and one at the second harmonic (SH) frequency.

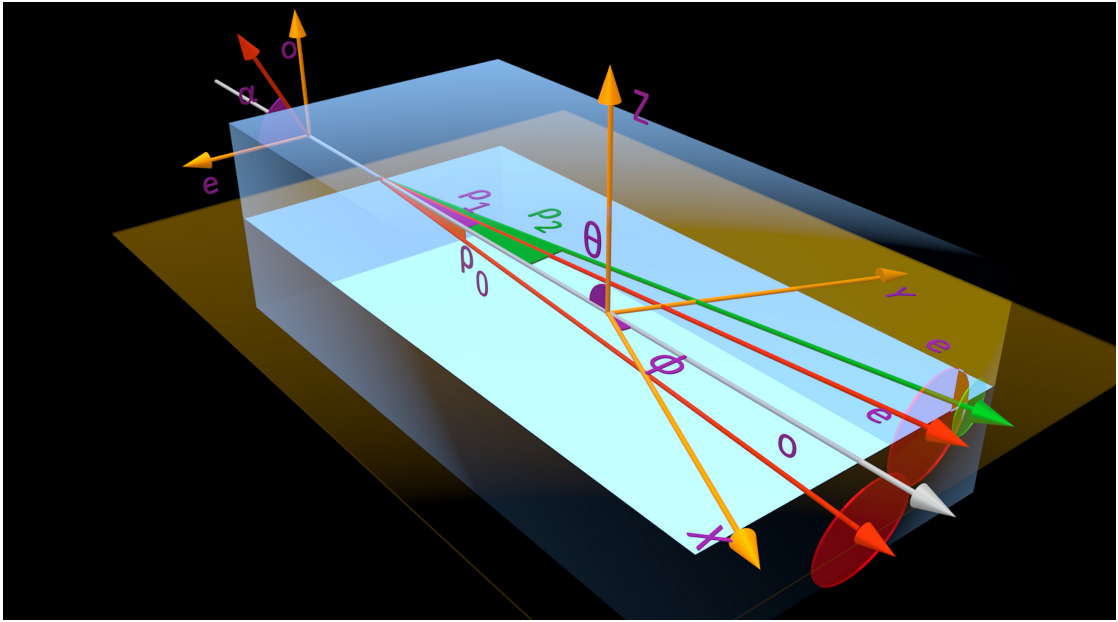

*Figure 1: graphical illustration of the beam propagation in the KTP crystals. The FF ordinary (o), FF extraordinary (e), and the SH beams have different walk-off angles, $\rho_0$, $\rho_1$, and $\rho_2$, respectively The orientation of the reference frame with respect to the crystallographic axes X, Y, Z is identified by the angles $\theta$ and $\varphi$. The orientation of the input linear polarization of the light beam is measured by the angle $\alpha$.*

Our experimental analysis was carried out in a sample of Potassium Titanyl Phosphate, a nonlinear crystal with quadratic susceptibility, commonly known as KTP. The crystal, 30 mm long, was cut to maximize the efficiency of second harmonic generation (SHG), by exploiting its anisotropic nature at room temperature. The nonlinear process is known as type-II SHG, as it involves the nonlinear coupling of the two polarization components of the input FF beam along the ordinary and extraordinary axis of the crystal and the SH wave, which is polarized along the extraordinary axis. Light propagation in anisotropic crystals (in the specific case, the KTP is an orthorhombic biaxial crystal) is quite peculiar, since the wavevector (the vector normal to the wave fronts) does not align, in general, with the direction of the energy flow (the Poynting vector). Figure 1 gives a visual interpretation of the process. The input optical beam is linearly polarized: the electric field forms an angle $\alpha$ with the horizontal reference axis. The white arrow represents the wavevector of the FF, while the Poynting vectors are indicated in red for the FF, and in green for the SH. The Poynting vector of the FF extraordinary wave (e) has a spatial walk-off angle $\rho_1$ with the horizontal axis. Similarly, the Poynting vector of the SH wave (e) has a spatial walk-off angle $\rho_2$. Figure 1 also shows the orientation of the KTP crystallographic axes, X,Y,Z, as well as the two relevant angles $\theta$ and $\varphi$ with respect to the propagation reference frame, which are used in order to control phase matching for type-II SHG.

Experiments have been carried close to the values $\theta = 90°$, $\varphi = 23.5°$, at the phase matching of the type-II SHG. Small angle tilts out of these reference values are a common way to introduce a controlled phase mismatch. In our experiments, we noticed the presence of a spatial asymmetry among the three waves, that we may explain by the presence of an additional walk-off of the (pseudo) ordinary axis, which we indicated by $\rho_0$. As it is well known, these walk-off angles and the consequent beam splitting induce limits the effective nonlinear interaction length along the propagation direction.

In order to demonstrate the generation of a spatial two-dimensional TSB, let us consider first a series of experimental results obtained with a polarization orientation angle $\alpha = 47°$: such a choice imposes a power unbalance in the FF between the ordinary and the extraordinary axis. Figure 2 summarizes the experimentally obtained spatial beam shapes of the FF and the SH, for different input intensities. When the intensity is relatively low, we observed only weak conversion into the SH, in absence of any change of shape in the FF (see Fig.2a for FF and 2e for SH). When we increased the input beam intensity up to 0.07 GW/cm$^2$, we observed the formation of a narrow coupled FF-SH beam, trapped at the intersection of the two diverging beams at the FF. Such self-focused beam, or TSB, has a diameter ~40 µm, that is about ten times smaller than the diameter of the input FF beam. The TSB, which results from a strong reshaping of the input FF beam, has the main features of a spatial soliton surrounded by radiation [27]: it involves both spectral components at the FF and the SH, as shown in Fig.2(b,c) and Fig.2(f,g). The maximum peak intensity of the TSB, which is 2.7 times higher than that of the encircling background, is reached for an input FF intensity of 0.64 GW/cm² (see Fig. 2(c)).

Now, surprisingly, we observed that by increasing further the input FF intensity, the TSB vanished (see Fig.2(d)): its existence is thus confined to a limited range of input intensity values. At relatively high intensities, the smooth FF background beam recovers a spatial profile which is nearly as wide as that observed at low intensities, except for the presence of minor residual distortions. To summarize, upon increase of the input laser intensity, the TSB is first formed, then it grows up till an intensity elevation which is up to nearly three times that of the local background, and then it disappears without leaving a visible trace. The observed TSB dynamics is thus very different from that of spatial solitons, which tend to persist and can even be strengthened upon increasing the input FF intensity. It is important to notice that as the TSB grows up, its shape is replicated at the SH, with a high contrast from the associated background. However, the beam dynamics at the FF and the SH are very different at very high intensities (cf. Fig.2(d) and Fig.2(h)).

As can be seen, although the high-energy located spot at the FF nearly disappears, beam breaking occurs at the SH. Specifically, for input intensities of 9.3 GW/cm², the SH breaks-up into a speckled beam (Fig. 2(h) and figure SM4 in the supplementary information). The process of beam-breaking at the SH has been studied in details in a related study [37].

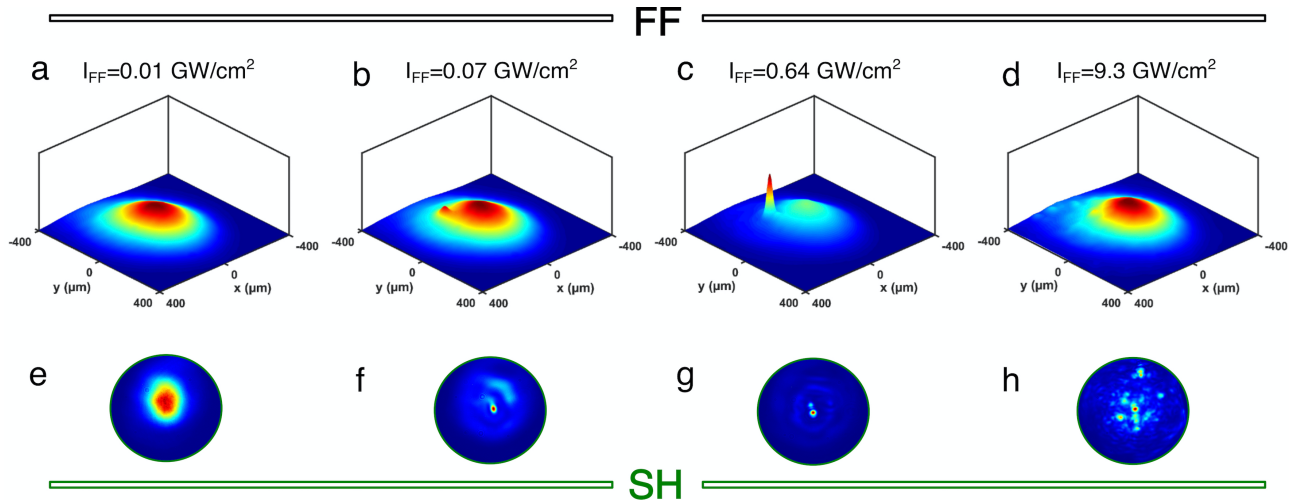

Figure 2: Output TSB profiles upon increasing input beam intensity $I_{FF}$. Panels **a,b,c,d** refer to the FF. Panels **e,f,g,h** show the corresponding SH intensity profile.

Figure 3 collects many different transverse beam sections at the FF vs. the input FF intensity: panel 3(a) clearly shows the appearance and the vanishing of the TSB. Whereas panel 3(b) clarifies the quantitative dependency of the TSB to background intensity contrast vs. the input intensity.

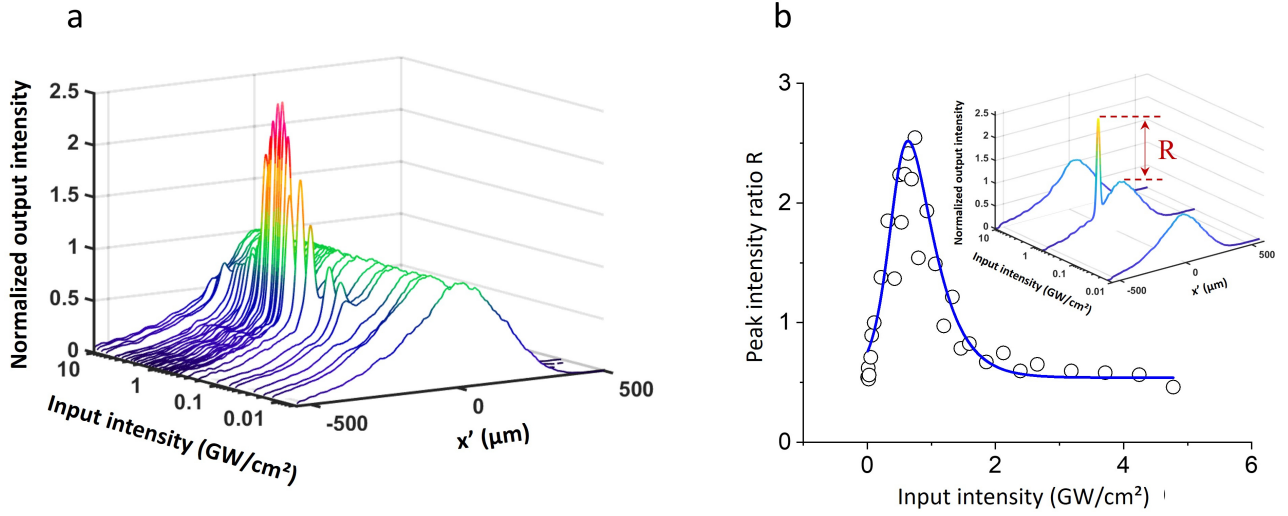

*Figure 3: **a** TSB sections of the FF component of the TSB, vs. input FF intensity; **b** peak intensity measuring the intensity elevation of the FF component of the TSB from the local background, vs. the input FF intensity.*

We observed that the TSB generation is influenced, similarly to quadratic spatial solitons, by the power unbalance between the two polarization components of the beam at FF, as determined by the polarization orientation angle α. Since the beams carried by each of two polarization components have slightly different propagation directions (owing to spatial walk-off), by modifying the polarization angle α, the spatial position of the TSB can abruptly hop from one output spatial position to another one. Although the TSB is systematically obtained for all angles from α = 10° to α = 80° for a fixed FF intensity, the elevation of the FF component of the TSB from the surrounding background and its spatial position varies: maximum TSB intensity is reached for two symmetric polarization orientation angles, close to either α = 30° and α = 60° (Fig. 4a) and the corresponding spatial shape is given in panel (4b). When the input polarization angle was set to α = 45° (i.e., equal input FF intensities in both polarization axes), we observed the presence of two competing TSBs (Figs. 4(c) and 4(d)). Panel (c) shows an example of spatial switching of the TSB position, for three particular values of α; while panel (d) illustrates the sharp nonlinear transition of the spatial TSB position for a set of values of the orientation angle α. Note that analogous dynamics were earlier reported for spatial solitons [38, 39]: the sharp dependence of the output beam position on the input state of polarization could find possible applications in ultrafast all-optical switching, as polarization comparator [24], or even, by extension, as an optical implementation of an activation function for an artificial neuron.

Additionally, in the large positive phase mismatch regime, we also observed the presence of multiple TSBs (see figure SM1 in the supplementary information).

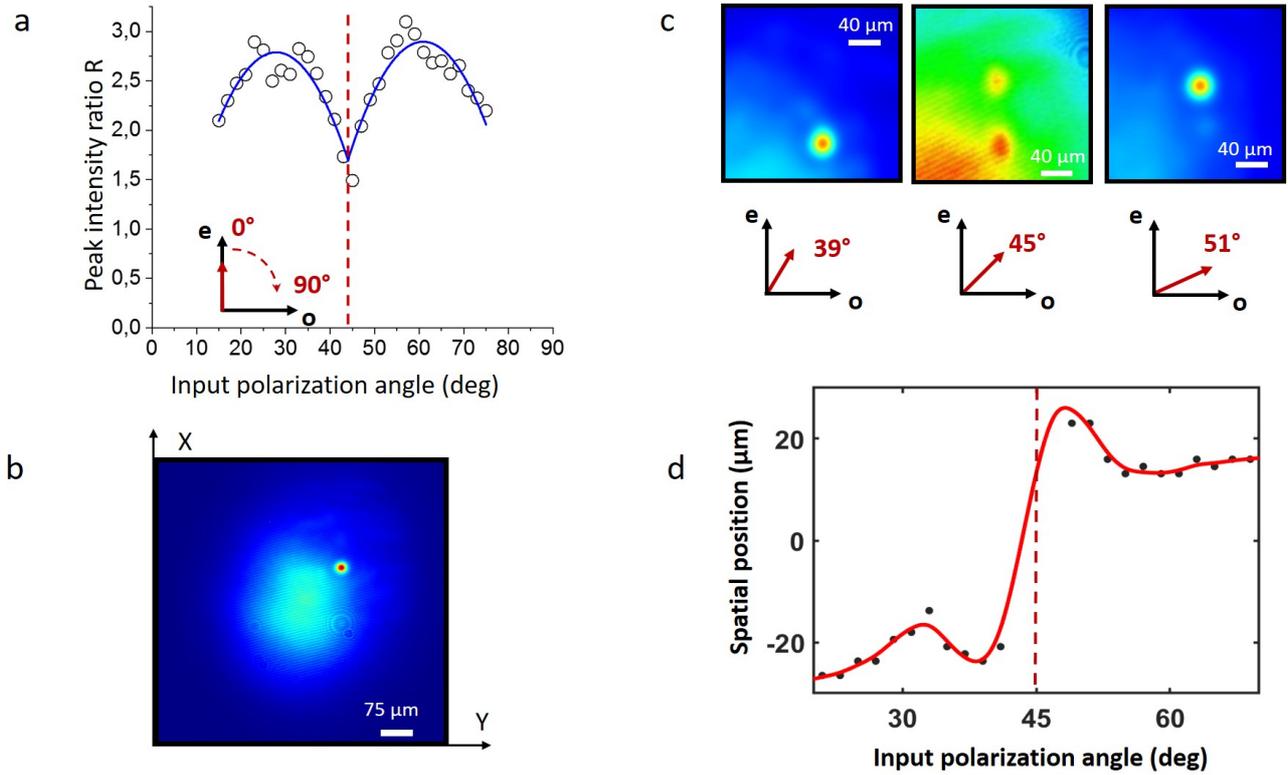

*Figure 4: Experimental results on the efficiency of the 2D spatial TSB trapping and associated spatial switching; **a** peak intensity ratio between the TSB and the input beam; **b** example of TSB generation for an input polarization orientation angle α = 60°; **c** example of TSB spatial switching for three different input polarization angles; **d** spatial position of the TSB vs. α ($I_{FF}$ = 0.64 GW/cm²).*

In our experimental analysis, we found that the generation of the TSB not only involves the spatial domain, as can be detected by a slow-time response camera, but it also has a direct influence on temporal shape of the FF component of the TSB. To analyze temporal reshaping effects in details, we measured the temporal autocorrelation within the TSB area, and we varied the input FF intensity (cf. Fig.5(a)) and the FF-SH phase mismatch. At low input intensities, the temporal autocorrelation trace has a Gaussian shape with a 41 ps temporal width (compatible with a pulse duration of 29 ps FWHMI), see Fig.5(b). However, once the TSB is formed, the shape of the autocorrelation trace takes a nearly triangular profile of 24 ps (FWHMI), see Fig.5(c). At even higher input powers, where the TSB is close to its vanishing, the temporal autocorrelation trace exhibits several peaks, which can be associated with a temporal break-up of the underlying pulse, see Fig.5(d). Interestingly, the same behavior, although with different intensities, is observed for both positive and negative values of the phase mismatch (cf. figures SM2 and SM3 in the supplementary information).

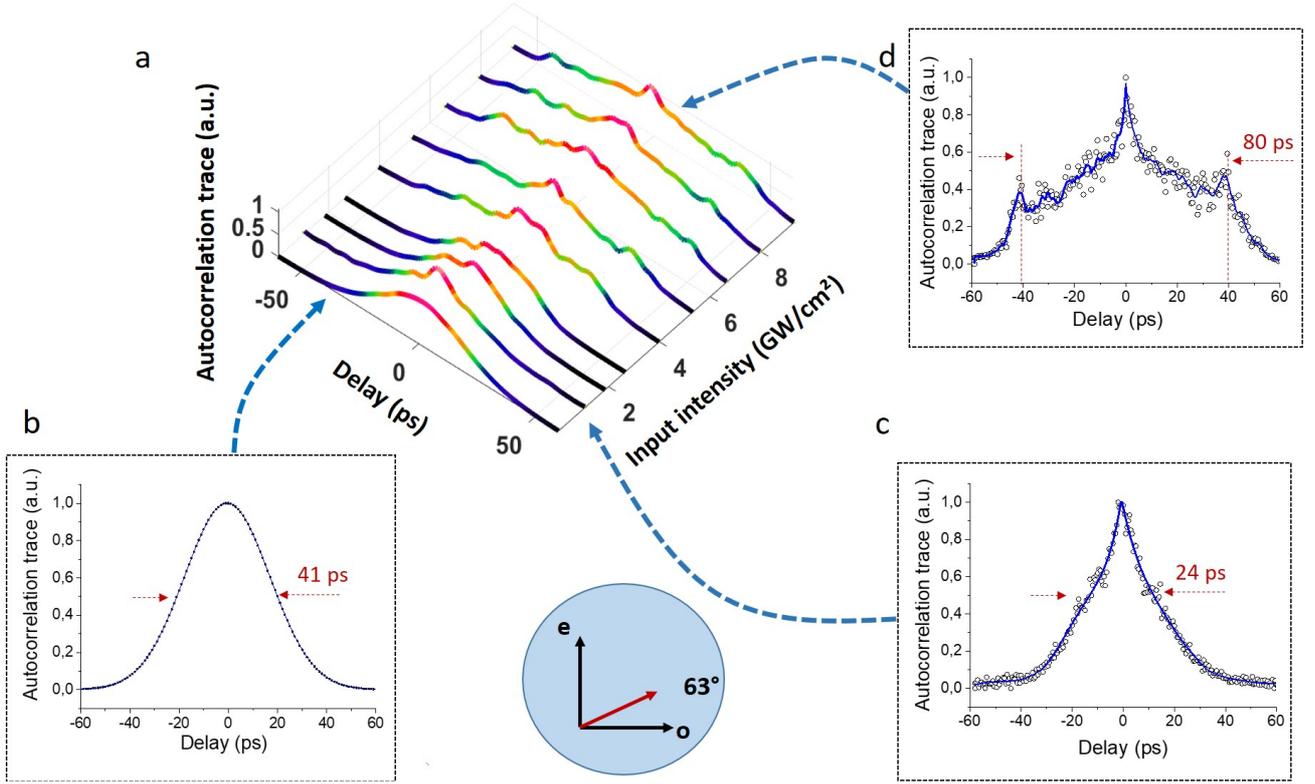

*Figure 5: Time domain analysis of pulses from the output FF component of the TSB.* ***a*** *autocorrelation traces at different input intensities, for perfect phase matching. Details of the traces:* ***b*** *autocorrelation of the input laser pulse;* ***c*** *autocorrelation at 1 GW/cm²;* ***d*** *autocorrelation taken at high FF intensities, such that the TSB has nearly vanished. The circular inset illustrates the orientation of the linear polarization state of the input beam.*

We would like to underline that the nature of the observed extreme TSB is deterministic and reproducible. We confirmed its deterministic nature by numerically solving the nonlinear coupled equations involving the three interacting waves, i.e., the two FF beams and the SH wave. In the numerical model, we included all relevant effects, namely diffraction and quadratic susceptibility. We limited our numerical analysis to the case of a simple spatial interaction of the three waves, thus neglecting the temporal dimension. Moreover, our analysis was limited to walk-off angles corresponding to the two extraordinary waves, and we set to zero, for simplicity, the FF ordinary beam angle $\rho_0$. The input beam had a Gaussian profile with 400 μm diameter, and we superimposed to it a weak bump, whose center had an offset of 200 μm in one of the spatial coordinates, which was used as a seed.

In figure 6, we show a collection of numerically computed output FF components of the TSB, for different input intensities. As can be seen by comparing with Fig.3a, numerical simulations agree well with experimental results. In particular, the intensity of the FF component of the TSB grows up to an elevation of about 2.5 times the background intensity. Then the TSB progressively disappears when the input FF beam intensity grows larger.

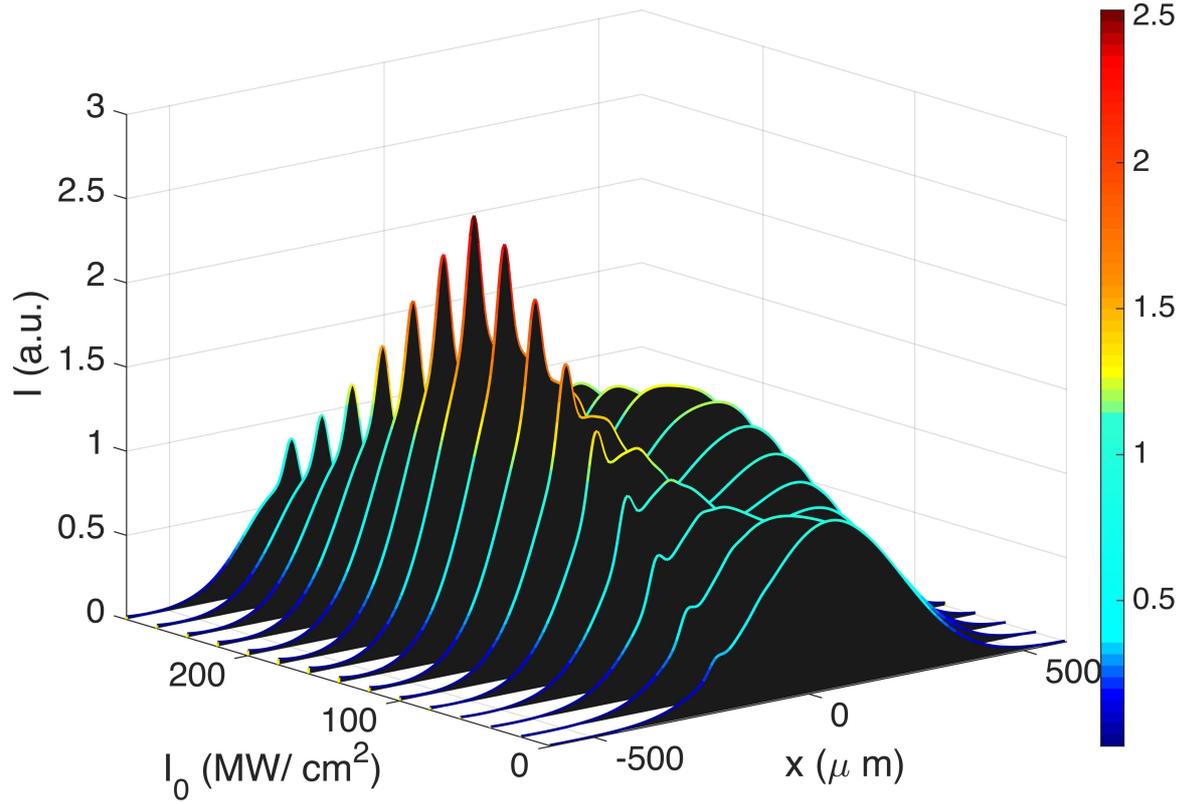

*Figure 6: Numerically simulated sections of the FF component of the TSB upon the input intensity $I_0$. The images have been collected after propagation of 30 mm.*

In conclusion, we demonstrated that a coupled FF and SH 2D spotlight beam can appear and disappear at the output of a quadratic crystal, in the presence of a monotonic increase of the input FF beam intensity. Such a beam bears a strong resemblance to fully localized waves such as the Peregrine temporal soliton of the 1D NLSE, which also appears and disappears as the input power grows larger, thanks to the equivalence between varying power or propagation length. However, the localized wave that we have unveiled here emerges in the frame of a genuine 2D nonlinear wave coupling process. Besides the resemblance with other types of extreme waves, the full nature of the extreme or rogue wave described in the present work appears to involve a coupling between its spatial and temporal dimensions. Additionally, we showed how the spatial position of the TSB can abruptly switch, upon a relatively small change of the linear polarization angle of the input FF beam. Although phase-mismatched SHG seems to be the dominant nonlinear process in the early stage of the TSB generation, the observed temporal wave breaking at the FF could provide the mechanism which is responsible for its disappearance at high intensities.

**Methods**
**Experiments.** We used a laser source at 1064 nm, delivering pulses of 30 ps with pulse energies up to several tens GW/cm² and a repetition rate of 10 Hz. The linearly polarized beam was collimated in the nonlinear crystal (beam diameter of 400 μm at 1/e² in intensity) by means of two convergent lenses and a half-wave plate, which allows for setting the polarization orientation of the input beam. An output convergent lens (f = 3.5 cm) was used to obtain a magnified image of the output end face of the crystal on a CCD camera. An autocorrelator was used to obtain the temporal evolution of the output FF beam. We used a type-II KTP crystal cut for second harmonic generation at 1064 nm. The spatial walk-off between the two FF beams and the SH is 3.48 mrad and 4.88 mrad, respectively. The phase matching conditions were controlled through crystal orientation in horizontal or vertical dimensions. Additional experimental results are discussed in the supplementary material. Temporal autocorrelations were obtained by selecting the TSB with a diaphragm, and then by measuring the autocorrelation at the FF.

**Numerical simulations.** We numerically solved the three-wave equations by considering the presence of diffraction, the spatial walk-off between the FF and SH, and the quadratic nonlinear response of the KTP at the relevant angles close to phase matching. The numerical integration scheme was based on a standard Runge-Kutta method. More information on the used equations and solver are given in reference [40].

**Acknowledgements**
We are grateful to A.B. Aceves, M. Ferraro, F. Mangini, Y. Sun, S. Trillo and M. Zitelli for helpful discussions. V.C. acknowledges the financial support provided by the French ANR through the "TRAFIC project: ANR-18-CE080016-01"; the CILAS Company (ArianeGroup) through the shared X-LAS laboratory; the "Région Nouvelle Aquitaine" through the projects F2MH; the National Research Agency under the Investments for the future program with the reference *ANR-10-LABX-0074-01 Sigma-LIM*. R. J and S. W. were supported by the European Research Council (ERC) under the European Union's Horizon 2020 research and innovation programme (No. 740355).

**Additional information**

**Competing financial interests**
The authors declare no competing financial interests.